# A wide-range temperature immune ultra-sensitive refractive index sensor using concatenated LPGs

**Saurabh Mani Tripathi[1], Arun Kumar[2], Wojtek J. Bock[1] and Predrag Mikulic[1]**

[1]*Photonics Research Center, Université du Québec en Outaouais, Québec, QC J8X 3X7, Canada.*

[2]*Physics Department, Indian Institute of Technology Delhi, New Delhi-110016, India.*

tripathi.iit@gmail.com, akumar@physics.iitd.ernet.in, wojtek.bock@uqo.ca, predrag.mikulic@uqo.ca

**Abstract:** Compensating the temperature induced phase-changes of concatenated dual-resonance long-period-gratings by a suitably chosen inter-grating material and space we report wide-range temperature insensitivity along with extremely high refractive-index sensitivity, on either side of turn-around wavelength.

1. **Introduction**

Temperature cross-sensitivity is one of the primary causes of errors associated with the fiber-optic sensors [1-3]. Being a fundamental material property, often a precise determination of refractive index (RI) needs temperature isolation/calibration, since the RI of samples (biological/chemical) and the waveguide regions changes with temperature. Recently we proposed and demonstrated a temperature insensitive RI sensor based on concatenated dual-resonance long period gratings (DRLPGs) with an appropriate inter-grating-space (IGS) in-between them [4] using a single (SMF-28$^{TM}$) fiber. Each LPG was tuned to the DRLPG regime by partially etching the cladding over the grating region, making the spectral variation of the phase-difference ($\Delta\beta = \beta_c - \beta_{cl}$), between the two modes involved, of opposite nature to that of the un-etched IGS region for $\lambda > \lambda_D$ ($\lambda_D$ is the turn-around wavelength for the DRLPG). As a result the resonance wavelengths over the higher wavelength side of the turn-around wavelength ($\lambda > \lambda_D$) became temperature insensitive for an appropriate length of IGS [4]. The required IGS length was quite high due to the same thermo-optic coefficients of the grating and IGS regions, increasing the overall sensor length.

In this paper we demonstrate a wide-range temperature insensitivity over the entire spectrum of DRLPG (on both the sides of $\lambda_D$) by choosing a $B_2O_3$ doped germanosilicate fiber as the IGS material. This is achieved due to the fact that the thermo-optic coefficients (dn/dt) of $B_2O_3$ (-3.5×10$^{-5}$ /°C) and $GeO_2$ (1.94×10$^{-5}$ /°C) are of the opposite nature. Further, the spectral variation of $\Delta\beta$ in the $B_2O_3$ doped germanosilicate fiber also shows a turnaround behavior (as shown later) even in the un-etched fiber used as IGS. The proposed scheme shows a considerable reduction in the overall sensor length and the wide-range of temperature insensitivity (over 1500-1700 nm), which is also suitable for various other applications like temperature-insensitive WDM channel isolation filter etc.

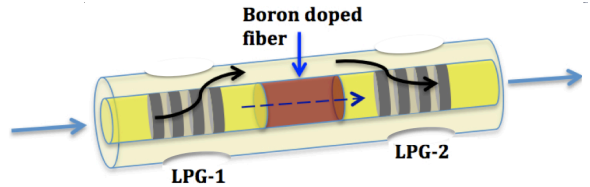

Fig.1: Schematic diagram of the sensor.

2. **Results and discussion**

In our experiments we first fabricated several DRLPGs in $GeO_2$ doped $SiO_2$ fiber (Corning SMF-28$^{TM}$) by partially etching the cladding [4]. Pair of LPGs with identical transmission spectrum were then selected and spliced axially with varying length of $B_2O_3$ doped germanosilicate fiber (Fibercore PS-1250/1500$^{TM}$) (see Fig.1). The typical measured RI and temperature sensitivities of the individual DRLPGs for the lower $\lambda_R$ (~1557 nm), are 1837 nm/RIU and 0.95 nm/°C, and for the upper $\lambda_R$ (~1627 nm) are 2464 nm/RIU and 1.03 nm/°C, respectively. The sensor was finally passed through a heating tube filled with water to measure its temperature response.

In Fig. 2(a) we plot the transmission spectrum of the sensor for an IGS = 8.9 mm, recorded at four different temperatures. Complete temperature insensitivity is evident over the entire wavelength range. The measured RI sensitivities of various transmission minima have been plotted in Fig.2(b), showing a maximum sensitivity of ~2577 nm/RIU obtained for the resonance minima near 1659 nm.

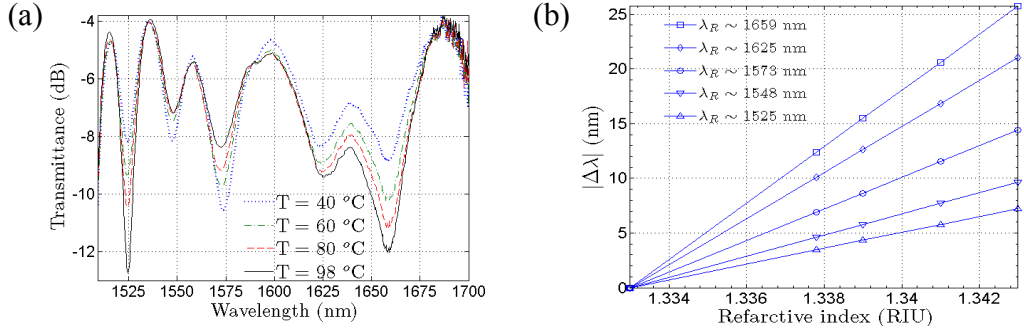

Fig.2: (a) transmission spectrum of the sensor for an IGS = 8.9 mm recorded at four different temperatures, (b) measured RI sensitivities corresponding to the resonance minima at 1525 nm, 1548 nm, 1573 nm, 1625 nm and 1659 nm.

In order to understand the temperature insensitivity theoretically, we consider the LPGs to be written in a 4.1 mole% $GeO_2$ doped $SiO_2$ (say Fiber-G) core [5] separated by an IGS with its core composition of 9.7 mole% $B_2O_3$ in 4.03 mole% $GeO_2$ and 86.27 mole% $SiO_2$ (say Fiber-B) [5]. The overall transmission of the sensor basically depends upon (*i*) the phase matching within the grating regions (used to excite and recouple cladding mode back to the core mode) and (*ii*) modal interference within the IGS region. Both of these factors are governed by the propagation constant difference ($\Delta\beta = \beta_c - \beta_{cl}$) between the core and cladding modes involved. In Fig. 3 we have plotted the spectral variation of $\Delta\beta$ for the above two fibers. It may be noted from Fig. 3(b) that the spectral variation of $\Delta\beta$ for Fiber-B also shows a turn around behavior. For a complete temperature insensitivity the $\lambda_D$ of the two curves must coincide, which has been obtained by taking the core and cladding radii as 4.1 μm and 55 μm [4] for fiber-G (Fig. 3(a)) and 9.9 μm and 62.5 μm for fiber-B (Fig. 3(b)). Two distinct features observed from these figures are that the temperature-induced variations to $\Delta\beta$ is (*i*) positive for Fiber-G and negative for Fiber-B, and (*ii*) much bigger for Fiber-B. A small length of IGS of Fiber-B can, therefore, compensate the thermal shifts introduced by the gratings in Fiber-G. This is clear from Fig.3 (c) and (d) where we have plotted the transmission spectrum of a single DRLPG in Fiber-G (of Fig. 3(a)) and two DRLPGs separated by 6.5 mm of Fiber-2 (of Fig. 3(b)), respectively. A complete temperature insensitivity on either side of $\lambda_D$ is evident in Fig. 3(d). The difference between the theoretical value of IGS (6.5 mm) to achieve temperature insensitivity with that of the experimental one (8.9 mm) can be attributed to the fact that the exact parameters of the fibers used in the experiment are not known and may be slightly different.

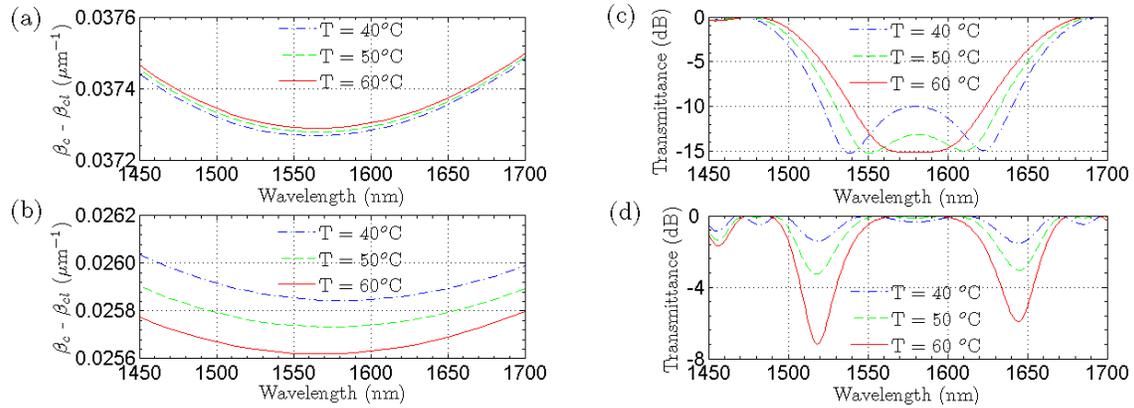

Fig.3: Spectral variation of Δβ for (a) Fiber-G and (b) Fiber-B. Theoretical transmission spectrum for (c) single DRLPG in Fiber-G and (d) concatenated DRLPGs in Fiber-G separated by 6.5 mm of Fiber-B. The ambient region is taken as water (RI= 1.333) throughout the calculations.